\documentclass[a4paper,12pt,fleqn]{article}

\usepackage{amsmath}

\newcommand{\pid}{\par\indent}
\newcommand{\pni}{\par\noindent}

\begin{document}

\title{Black-body laws derived from \\ a minimum knowledge of Physics}
\author{ A.G.Agnese${}^{1}$, M.La Camera${}^{2}$ 
and E.Recami${}^{3}$}
\date{}
\maketitle 
\begin{center}
${}^{1,2}${\em Dipartimento di Fisica dell'Universit\`a di Genova\\
Istituto Nazionale di Fisica Nucleare,Sezione di Genova\\
Via Dodecaneso 33, 16146 Genova, Italy}\\
${}^{3}$\em{Facolt\`a di Ingegneria dell'Universit\`a di Bergamo\\
Via Marconi 5, 24044 Dalmine (BG), Italy\\
Istituto Nazionale di Fisica Nucleare,Sezione di Milano}\\
\end{center} \bigskip
\begin{abstract}
Starting from the knowledge of the four fundamental quantities 
length $L$, mass $M$, time $T$, absolute temperature $\theta$  and 
accepting the validity of Gauss's law in all dimensions, we 
generalize, by the theory of physical dimensions, the 
expressions of the Stephan-Boltzmann law and of the Planck's 
formula for the black-body radiation to a space-time with one time 
and $n$ space coordinates. In the particular case $n=3$ we shall 
recover the known results.
\end{abstract} \bigskip \pni
 \vspace{1in}\pni
$\overline{\hspace{2in}}$\pni
${}^{1}$E-mail: agnese@ge.infn.it\pni
${}^{2}$E-mail: lacamera@ge.infn.it\pni
${}^{3}$E-mail: erasmo.recami@mi.infn.it
\newpage

\baselineskip = 2\baselineskip

\section{Introduction} 

Applications of dimensional analysis to  mechanical problems are
well known in physics. \pid
From a historical point of view, the first seeds of that method can be 
found in Newton's ``\emph{Principia}'' ${}^{1}$ and in Fourier's 
``\emph{Th\'eorie Analytique de la Chaleur}'' ${}^{2}$. 
Successively, important contributions to the modern treatment of 
dimensional theory have been made by R.C.Tolman who defined the 
principle of dimensional homogeneity and stated the $\Pi$ 
theorem ${}^{3}$. \pid
It is worth noticing that dimensional analysis was employed by 
Albert Einstein too in the calculus of the eigenfrequencies in 
solid bodies  ${}^{4}$. He set forth the pendulum example and, on 
obtaining $T= C \sqrt{\dfrac {l}{g}} $ as usual, said: ``\emph{One can, 
as is known, get a little more out from dimensional considerations, 
but not with complete rigor. The dimensionless numerical factor 
(like $C$ here), whose magnitude is only given by a more or less 
detailed mathematical theory, are usually of the order of unity. We 
cannot require this rigorously, for why shouldn't a numerical factor 
like $(12\pi )^{3}$ appear in a mathemathical-physical deduction? But 
without doubt such cases are rarities \ldots }''  ${}^{5}$.\pid
For a more complete view of the subject, we refer the reader to the 
book of Bridgman~${}^{6}$.\pid
Our aim here is deducing the form of the Stephan-Boltzman law and of 
the Planck's formula in a  space-time with one time and $n$ space 
coordinates, by employing the methods of dimensional analysis only. 
The minimum knowledge of physics we need getting started is given by 
the definitions of length $L$, mass $M$, time $T$ and absolute 
temperature $\theta$; we also accept the validity of Gauss's law  in 
all dimensions and know the definitions of force and energy. Finally, 
in the particular case $n=3$ we shall recover the known results
on the basis of little physical information.\pid

\section{The universal constants and the $\Pi$ theorem}

Let us recall here some elements of physical dimensions theory. 
For instance, if we aim at a physical theory of gravity, the three 
fundamental units $L,M$ and $T$ are sufficient, but they must be 
associated as usual to some universal constants. In the gravitational 
case two universal constants are needed: an invariant velocity $c$ with 
dimensions $[c] = [LT^{-\, 1}]$ and a gravitation constant $G_{n}$, 
where the subscript $n$ stands for the number of space coordinates, 
with physical dimensions $[G_{n}] = [L^{n}M^{-\, 1}T^{-\, 2}]$. The 
latter dimensional equality follows immediately from Gauss's law for 
the gravitational field $\boldsymbol{g}_{n}$ produced by a 
$n$-dimensional distribution of mass $\rho_{n}$:
\begin{equation}
\boldsymbol{\nabla \cdot g}_{n} = -\, \dfrac{2\pi^{n/2}}
{\Gamma (n/2)}\, G_{n}\rho_{n}
\end{equation}
and from the relation
\begin{equation}
\boldsymbol{F}_{n} =  m \boldsymbol{g}_{n} 
\end{equation}
which gives the force $\boldsymbol{F}_{n}$ (with dimensions 
$[LMT^{-\, 2}]$) acting on a test mass $m$ located in the gravitational 
field $\boldsymbol{g}_{n}$.\pid
If the source is a point particle of mass $M$ we have, in a 
$n$-dimensional space:
\begin{equation} |\boldsymbol{g}_{n}| = 
\begin{cases} 
\dfrac{G_{n}M}{r^{n-1}} \hspace{2in} &\text{for} \quad n>1 \\ {}\\
\dfrac{G_{1}M}{2} \hspace{2in} &\text{for} \quad n=1
\end{cases}\end{equation}
In dealing with electromagnetism, we need again $c$ and another 
fundamental constant.We do not introduce any additional physical 
quantities besides $L,M$ and $T$; so, 
Gauss's law for the electrostatic field $\boldsymbol{E}_{n}$ 
produced by a $n$-dimensional distribution of charge $\rho_{n}$ will
be written, in Gaussian units, as:
\begin{equation}
\boldsymbol{\nabla \cdot E}_{n} = \dfrac{2\pi^{n/2}}{\Gamma (n/2)}\, 
\rho_{n}
\end{equation}
Acting as in the gravitational case, the electrostatic field 
$\boldsymbol{E}_{n}$ of a point charge $Q_{n}$ (whose dimensions 
turn out to be $[L^{n/2}M^{1/2}T^{-\, 1}]$) is : 
\begin{equation} |\boldsymbol{E}_{n}| = 
\begin{cases} 
\dfrac{Q_{n}}{r^{n-1}} \hspace{2in} &\text{for} \quad n>1 \\ {}\\
\dfrac{Q_{1}}{2} \hspace{2in} &\text{for} \quad n=1
\end{cases}\end{equation}
Before going on, we notice that in the three dimensional space 
Coulomb's law for two point charges $Z_{1}e$ and $Z_{2}e$
\begin{equation}
|\boldsymbol{F}| = \dfrac {Z_{1}Z_{2}e^{2}}{r^{2}}
\end{equation}
can be written, by introducing into it the fine structure 
constant\linebreak $\alpha = e^{2}/(\hbar c)$, as 
\begin{equation}
|\boldsymbol{F}| = \dfrac {Z_{1}Z_{2}\alpha\hbar c}{r^{2}}
\end{equation}
So the other fundamental constant we need can be either the electron 
charge $e$ or the Planck's constant $\hbar$. Therefore when dealing
within both gravity and electromagnetism, we can associate $(L,M,T)$  
with $(c,G,e)$ or with $(c,G,\hbar )$: this agrees with the choices
made, respectively by Stoney in $1874$ ${}^{7}$ and by Planck in 
$1899$ ${}^{8}$. In our approach the appearance of the Planck's 
constant $\hbar$ is clearly related to the quantization of the electric 
charge, however.\pid
Let us recall that, according to Planck, ${}^{8}$ the system of 
``natural units'' must include $\sqrt{(\hbar c)/G}$ as unit of mass, 
$\sqrt{(\hbar G)/c^{3}}$ as unit of length and $\sqrt{(\hbar G)/c^{5}}$  
as unit of time; while according to Stoney ${}^{7}$ the ``natural 
units'' for mass, length and time are $e/\sqrt{G}$, 
$(\sqrt{G} e)/c^{2}$ and $(\sqrt{G}e)/c^{3}$ respectively. 
One sees immediately that Stoney's units are 
given by Planck's  times the square root of the fine structure 
constant. By the way, one may notice that Stoney's conception of the 
electron is somewhat different from that resulting from the subsequent 
discovery by Thomson.\pid
Going back to our task, it is suitable, in $n$ spatial dimensions, 
to pass from $G$ to $G_{n}$ and from $\alpha$ to 
$\alpha_{n} = e_{n}^{2}/(\hbar_{n} c)$; so, being $\alpha_{n}$ 
dimensionless by definition, the replacement of $e$ with $e_{n}$ and 
of $\hbar$ with $\hbar_{n}$ makes their respective dimensions to be 
$[e_{n}] = [L^{n/2}M^{1/2}T^{-\, 1}]$ and 
$[\hbar_{n}] = [L^{n-\, 1}MT^{-\, 1}]$. Of course, 
the quantities labelled with $n$ are defined only formally and not 
operationally, but this procedure will provide interesting results.\pid
Let us fix our attention at this point on ordinary thermodynamics. 
In this case we need a fourth  quantity, namely, the absolute 
temperature $\theta$ which then must be associated with another 
universal constant. In a quite natural way, we choose the Boltzmann 
constant $k$ whose dimensions are $[L^{2}MT^{-\, 2}\theta^{-\, 1}]$.
\pid Consequently in the dimensional equations which refer to the 
thermodynamics of the electromagnetic radiation  factors of the type 
$c^{\alpha}\hbar_n^{\beta}k^{\gamma}$ must appear.\pid
Let us finally recall the so called $\Pi$ theorem ${}^{6}$ which we
state as follows: ``Denote by $P_{1},P_{2},\ldots P_{r}$ the 
magnitudes of quantities which may be physical magnitudes or 
experimental constants. Suppose that a functional relation 
$f(P_{1},P_{2},\ldots P_{r}) = 0$ holds, 
and furthermore that this is a complete equation 
and is the only relation between the magnitudes. Then, if there are 
$s$ fundamental magnitudes the relation can be expressed in the form 
$F(\Pi_{1},\Pi_{2},\ldots \Pi_{r-\, s}) = 0$ where the $\Pi$'s are 
the $r-\, s$ independent products of the arguments $P_{1},P_{2},
\ldots P_{r}$ which are dimensionless in the fundamental magnitudes''.
  
\section{The Stefan-Boltzmann law}

Now we are ready to establish the Stefan-Boltzmann law in a 
$n$-dimensional space. Denoting by $u_{n}(\theta )$ the energy 
density of the electromagnetic radiation enclosed in a cavity and
in equilibrium at the temperature $\theta$, the 
functional relation quoted in the $\Pi$ theorem is $f(u_{n},\theta ,c,
\hbar_{n},k) = 0$. In this case $r-\, s=5-\, 4=1$, so there is only 
one product $\Pi$ which can be written in the form:
\begin{equation}
\Pi_{1} = u_{n}^{-1} c^{\alpha_1}
\hbar_{n}^{\beta_1}k^{\gamma_1} \theta^{\delta_1} 
\end{equation} 
Remembering that $[u_{n}] = [L^{2-n}MT^{-2}]$ and that consequently 
we shall have to discard the values $n=1$ and $n=2$ by physical 
reasons, we obtain the algebraic system:
\begin{equation}
\left\{ \begin{split} &\beta_1 + \gamma_1 - 1 = 0  \\
&\alpha_1 + (n-1)\beta_1 + 2\gamma_1 + n -2 = 0 \\
&\alpha_1 + \beta_1 + 2\gamma_1 -2 = 0  \\ &\gamma_1 -
\delta_1 =0 \end{split} \right.
\end{equation}
For $n \geq 3$ the solutions are:
\begin{equation}
\left\{ \begin{split} &\alpha_1 = -\, \dfrac {n}{n-2} 
\\ &\beta_1 = -\, \dfrac {n}{n-2}  \\ &\gamma_1 =\dfrac
{2(n-1)}{n-2}  \\ &\delta_1 = \dfrac {2(n-1)}{n-2}
\end{split} \right.
\end{equation}
and therefore
\begin{equation}
u_n \propto \, \dfrac {(k\theta)^{2(n-1)/(n-2)}} 
{(\hbar_n c)^{n/(n-2)}}
\end{equation}
When $n=3$ we recover the Stephan-Boltzmann law:
\begin{equation} 
u \propto \, \dfrac{(k\theta)^4}{(\hbar c)^3}
\end{equation} 
Obviously the numerical factor $\pi^{2}/15$ which appears in that 
law cannot be determined by dimensional analysis.
\section{The Planck's formula}

Let us now obtain the Planck's formula in a $n$-dimensional space.\pid
Calling $u_{n,\omega}(\omega ,\theta )$ the spectral energy density 
of the black-body radiation in the frequency interval 
between $\omega$ and $\omega + d\omega$, the functional relation is 
now $f(u_{n,\omega },\omega ,\theta ,c,\hbar,k)=0$. In this case
$r-s = 6-4 = 2$, so there are two products $\Pi$'s which will be
written as
\begin{align}
\Pi_2 &= u_{n,\omega }^{-1}c^{\alpha_2}\hbar^{\beta_2}k^{\gamma_2}
\omega^{\delta_2} \\ 
\Pi_3 &=\theta^{-1}c^{\alpha_3}\hbar^{\beta_3}k^{\gamma_3}
\omega^{\delta_3} 
\end{align}
and satisfy a relation of the form $F(\Pi_2,\Pi_3) = 0$. \pid
The two algebraic systems one obtains are:
\begin{equation} \hspace{-0.5in}
\left\{ \begin{split} &\beta_2 + \gamma_2 - 1 =0  \\
&\alpha_2 + (n-1)\beta_2 + 2\gamma_2 + n - 2 = 0  \\
&\alpha_2 + \beta_2 + 2\gamma_2 + \delta_2 - 1 = 0  \\
&\gamma_2 = 0  \end{split} \right. 
\hspace{0.3in}
\left\{ \begin{split} &\beta_3 + \gamma_3 = 0  \\
&\alpha_3 + (n-1)\beta_3 + 2\gamma_3 = 0 \\ &\alpha_3 +
\beta_3 + 2\gamma_3 + \delta_3 = 0  \\ &\gamma_3 + 1 = 0
\end{split} \right.
\end{equation}
with the respective solutions:
\begin{equation} \hspace{-0.5in}
\left\{ \begin{split} &\alpha_2 = -\, (2n - 3)  \\
&\beta_2 = 1  \\ &\gamma_2 = 0  \\ &\delta_2 = 2n - 3
\end{split} \right. 
\hspace{1.5in}
\left\{ \begin{split} &\alpha_3 = -\, (n - 3)  \\
&\beta_3 = 1  \\ &\gamma_3 = -\, 1  \\ &\delta_3 = n - 2
\end{split} \right.
\end{equation}
Therefore from the relation $F(\Pi_2,\Pi_3) = 0$ it follows that
\begin{equation}
u_{n,\omega} = \dfrac {\hbar_n \omega^{2n-3}}{c^{2n-3}}\, \varphi
\left( \dfrac {\hbar_n \omega^{n-2}}{c^{n-3}k\theta}\right)
\end{equation}
On the other hand it results, by definition, that $\int_{0}^{\infty} 
u_{n,\omega}(\omega ,\theta )\, d\omega = u_n(\theta )$; so, by using 
the expressions already found for $u_{n,\omega}(\omega ,\theta )$ and 
for $u_n(\theta )$, we obtain the following equation to be 
satisfied by the unknown function $\varphi$:
\begin{equation}
\int_0^\infty \dfrac {\hbar_n \omega^{2n-3}}{c^{2n-3}}\, \varphi
\left( \dfrac {\hbar_n \omega^{n-2}}{c^{n-3}k\theta}\right) 
d\omega \propto \dfrac {(k\theta)^{2(n-1)/(n-2)}}
{(\hbar_n c)^{n/(n-2)}} 
\end{equation}
Passing to the new variable  $\varepsilon_n = (\hbar_n
\omega^{n-2})/c^{n-3}$, whose meaning as a photon energy is apparent, 
the last equation becomes: 
\begin{equation}
\dfrac {1}{(n-\, 2)}\,\int_0^\infty 
\varepsilon_n^{2(n-1)/(n-2)\,-\, 1} \varphi \left(\dfrac 
{\varepsilon_n}{k\theta}\right)\, d\varepsilon_n  \propto 
(k\theta)^{2(n-1)/(n-2)}
\end{equation} 
As to the function $\varphi$, some considerations are in order.\pid 
First of all $\varphi$ cannot contain monomial terms in 
$(\hbar_n \omega^{n-2})/(c^{n-3}k\theta)$, otherwise it would  
modify the correct depedendence in Eq.(17) of $u_{n,\omega}$ on the 
term $\omega^{2n-\, 3}$ which comes by multiplying the 
factor $\omega^{n-\, 2}$ contained in the photon energy 
$\varepsilon_n$ by the factor $\omega^{n-\, 1}$ due to the evaluation 
of the number of photon states in the interval $d\omega$.\pid
Second, when $n=3$ we know the behavior of 
$u_{n,\omega}$ in the limits $\omega \to 0$ (Rayleigh-Jeans's law) 
and $\omega \to \infty$ (Wien's law), so, when $n=3$, one must have
\begin{equation} 
\varphi \left( \dfrac{\hbar \omega }{k\theta } \right) \sim 
\begin{cases}
\left( \dfrac {\hbar\omega }{k\theta }\right)^{-\, 1} \hspace{1.25in} 
{\text for} \quad \omega \to 0  \\ {} \\
\exp\left(-\,\dfrac {\hbar \omega }{k\theta }\right) \hspace{1in} 
{\text for} \quad \omega \to \infty
\end{cases}\end{equation} 
From a strictly mathematical point of view, the function $\varphi$ 
which satisfies Eq.(19) is not unique, but, taking into account our 
requirements, the choice is very restricted and, discarding too 
involved expressions, one is left only with the function
\begin{equation}
\varphi \left(\dfrac {\varepsilon_n}{k\theta }\right) \,\propto
\,\dfrac {1}{\exp{\left(\dfrac {\varepsilon_n}{k\theta }\right)}- 1}    
\end{equation}
Once again, we notice that for $n=3$ Eq.(17) becomes
\begin{equation}
u_{n,\omega} \propto \dfrac {\hbar_n \omega^{2n-3}}{c^{2n-3}}\,
\dfrac {1}{\exp{\left(\dfrac {\varepsilon_n}{k\theta }\right)}- 1}    
\end{equation} 
so that one recovers, apart from a numerical factor $1/\pi^{2}$, the 
Planck's formula and hence the Bose-Einstein statistics.\pid 
Had we ignored the behavior of the function $\varphi$ as $\omega 
\to 0$, the choice made in Eq.(21) might have been changed to $\varphi 
(\varepsilon_{n}/(k\theta )) \propto 
\exp (-\,\varepsilon_{n}/(k\theta ))$, which is proper for the
Boltzmann statistics, or even to $\varphi (\varepsilon_{n}/(k\theta ))
\propto (\exp (\varepsilon_{n}/(k\theta )) + 1)^{-\, 1}$, which is 
proper for the Fermi-Dirac statistics.

\section{Conclusions}

We would like to focus attention on some interesting properties of 
dimensional analysis we made use of in this 
paper.\pid 
First of all, in treating by the dimensional method a quite general 
problem of physics, it appeared  of capital importance to associate 
the four fundamental quantities ($L,M,T,\theta $) with as 
many universal constants, namely ($c,G_{n},\hbar_{n},k$). In less 
general situations, one needs only some of these costants; for example, 
when restricting to the three-dimensional space, one can select $c$ 
and $G$ for treating gravity, $c$ and $\hbar$ (or $e$) for treating 
electromagnetism or, as in our case, $c,\hbar$ and $k$ for treating 
the electromagnetic radiation thermodynamics.\pid
On the other hand, such a prescription was alredy contained in the 
works of Stoney ${}^{7}$ and of Planck ${}^{8}$. In Planck's words 
natural units are such because\linebreak ``\emph{\ldots they are
independent of particular bodies or substances, necessarily retain 
their meanings in all times and in every culture even if 
extra-terrestrial or extra-human.}'' ${}^{9}$ \pid
Secondly, let us comment on the power of the physical dimensions 
theory to test the soundness of a physical law. 
To make an example let us consider the Rayleigh-Jeans's formula for 
black-body radiation. We can obtain it if, contrary to what we just 
stated, we  accept the  widespread belief that electromagnetism 
can be treated by employing as universal constant only 
$c$, disregarding $\hbar$ (or $e$). Acting as before, we would write, 
in three space dimensions, the new functional relation among the 
quantities of interest as 
$f(u_{,\omega},\omega ,\theta ,c,k) = 0$. Now we have $r-s = 5 - 4 =1$ 
so there is only the following product
\begin{equation}
\Pi = u_{,\omega}^{-\, 1} \omega^{\alpha} \theta^{\beta} c^{\gamma} 
k^{\delta}
\end{equation} 
The algebraic system is
\begin{equation}
\left\{ \begin{split} &\alpha +\gamma + 2\delta - 1 = 0  \\
&\beta - \delta = 0 \\
&\gamma + 2\delta + 1 = 0  \\ &\delta - 1 = 0
\end{split} \right.
\end{equation}
and its solutions are:
\begin{equation}
\left\{ \begin{split} &\alpha = 2\\ &\beta = 1 \\ &\gamma = -\, 3\\
&\delta = 1
\end{split} \right.
\end{equation}
The final result
\begin{equation}
u_{,\omega}(\theta ) \propto (k\theta \omega^{2})/c^{3}
\end{equation}
reproduces indeed, apart from numerical factors, the Rayleigh-Jeans's 
formula. As to the Stephan-Boltzmann law one easily realizes that, 
when removing $\hbar$, Eq.(8) leads to $r-s = 4-4 = 0$ and hence no 
physical law is obtainable, in accordance with the fact that now the 
energy density $u(\theta )$ blows up to infinity.\pid
Lastly, we wish to remind the reader that a gravitational constant 
$G_{n}$ with dimensions $L^{n}M^{-\, 1}T^{-\, 2}$ is currently used in 
higher dimensional theories of gravity; so, it seems quite consistent 
that other universal constants such as $e$ or $\hbar$ be 
extended to $n$ dimensions! One must take however into account the 
following. We have seen that the photon energy is
$\varepsilon_{n} = \hbar_{n}\omega^{n-2}/c^{n-\, 3}$ and hence its
momentum is $|\boldsymbol{p}_{n}| =\varepsilon_{n}/c = \hbar_{n} 
(\omega /c)^{n-2} = \hbar_{n} |\boldsymbol{k}|^{n-2}$; if now we pass 
to  matter waves and make use of the de Broglie's relations, it 
happens that only for $n = 3$ the connection is relativistically 
invariant.\pid
Finally, we would like to add a brief comment about the
so-called ``black-hole thermodynamics'', invented by authors so as 
Bekenstein ${}^{10}$ and Hawking~${}^{11}$ by generalizing the 
ordinary thermodynamics.\pid
For instance, a black-body absorbs by definition every 
electromagnetic radiation impinging on it, and was shown to emit a 
characteristic (Planck) electromagnetic spectrum. By contrast, a 
black-hole is defined as a body absorbing every kind of particle and 
radiation impinging on it, and is predicted to emit a characteristic 
spectrum for each kind of particle or radiation. Such spectra are 
expected to be Planck-like, except that in their denominator the 
minus sign (valid in general for boson emissions) has to be replaced 
by a plus sign in the case of fermion emissions.\pid
These beautiful results have been reached by a clever, simultaneous 
recourse to general relativity, quantum mechanics and thermodynamics. 
They still lack, therefore, a definite theoretical foundation. From 
this point of view, it may be interesting that they get some support 
by simple dimensional considerations. 

\begin{flushleft}
{\large{\textbf{Acknowledgments}}}\\ \vspace{0.2in}
We are grateful to Massimiliano Badino, Decio Pescetti and 
Nadia Robotti for their comments and suggestions.
\end{flushleft}


\newpage

\begin{thebibliography}{00}
\bibitem{1} I.Newton, \emph{Philosophiae Naturalis Principia 
Mathematica}, Liber Secundus, ``De motu corporum'' Prop. XXX (1687)
\bibitem{2} J.B.J.Fourier, \emph{Th\'eorie Analytique de la 
Chaleur}, Section IX, Art.157-162 (1822)
\bibitem{3} R.C.Tolman, ``The principle of similitude and the principle 
of dimensional homogeneity,'' Phys.Rev. \textbf{6}, 219-231 (1915)
\bibitem{4} A. Einstein, ``Elementare Betrachtungen \"{u}ber die 
termische Molecular bewegung in festen K\"{o}rpen,'' Ann.d.Phys. 
\textbf{35}, 679-694 (1911)
\bibitem{5} Translated into English by E.U.Condon. ``Where do we live? 
Reflections on physical Units and the Universal Constants,'' Am.Jour.of 
Phys. \textbf{2}, 63-69 (1934)
\bibitem{6} P.W.Bridgmann, ``\emph{Dimensional Analysis}'' Yale 
University Press; issued as a Yale Paperbond, January 1963 
\bibitem{7} G.J.Stoney, ``On the Physical Units of Nature,'' 
Phil.Mag. \textbf{11}, 381-391 (1881)
\bibitem{8} M.Planck, ``\"{U}ber irreversible 
Strahlumgsvorg\"{a}uge.5 Mitteilung,''\\ S.-B.Preu\ss .Acad.Wiss. 
\textbf{5}, 440-480 (1899)
\bibitem{9} Our tranlation from ref.(8)
\bibitem{10} J.D.Bekenstein, ``Generalized second law of 
thermodynamics in black-hole physics,'' Phys.Rev. \textbf{D9}, 
3292-3300 (1974)
\bibitem{11} S.W.Hawking, ``Particle creation by black-holes,'' 
Commun.Math.Phys. \textbf{43}, 199-220 (1975)

\end{thebibliography}
\end{document}